\documentclass[]{emulateapj}
\usepackage[backref, breaklinks, colorlinks, citecolor=blue, linkcolor=magenta]{hyperref}
\usepackage[dvipsnames]{xcolor}
\usepackage{times}
\usepackage{amsmath}

\slugcomment{Submitted to The Astrophysical Journal}

\shorttitle{One-Two Quench}
\shortauthors{Sanchez et. al.}

\usepackage{natbib}
\begin{document} 

\title{One-Two Quench: A Double Minor Merger Scenario}

\author{N. Nicole Sanchez\altaffilmark{1}}
\author{Michael Tremmel\altaffilmark{2}}
\author{Jessica K. Werk\altaffilmark{1}}
\author{Andrew Pontzen\altaffilmark{3}}
\author{Charlotte Christensen\altaffilmark{4}}
\author{Thomas Quinn\altaffilmark{1}}
\author{Sarah Loebman\altaffilmark{5}}
\author{Akaxia Cruz\altaffilmark{6}}

\affil{$^1$Astronomy Department, University of Washington, Seattle, WA 98195, US, sanchenn@uw.edu}
\affil{$^2$Yale Center for Astronomy \& Astrophysics, Physics Department, P.O. Box 208120, New Haven, CT 06520, USA}
\affil{$^3$Department of Physics \& Astronomy, University College London, 132 Hampstead Road, London, NWI 2PS, United Kingdom}
\affil{$^4$Physics Department, Grinnell College, 1116 Eighth Ave., Grinnell, IA 50112, United States}
\affil{$^5$Department of Physics, University of California, Davis, CA 95616, US}
\affil{$^6$Physics Department, University of Washington, Seattle, WA 98195, US}

\begin{abstract}\label{abs:abstractlabel}
 
Using the N-body+Smoothed particle hydrodynamics code, ChaNGa, we identify two merger-driven processes\textemdash disk disruption and supermassive black hole (SMBH) feedback\textemdash which work together to quench L$^*$ galaxies for over 7 Gyr. Specifically, we examine the cessation of star formation in a simulated Milky Way (MW) analog, driven by an interaction with two minor satellites. Both interactions occur within $\sim$100 Myr of each other, and the satellites both have masses 5 to 20 times smaller than that of their MW-like host galaxy. Using the genetic modification process of \cite{Roth2016}, we generate a set of four zoom-in, MW-mass galaxies all of which exhibit unique star formation histories due to small changes to their assembly histories. In two of these four cases, the galaxy is quenched by $z = 1$. Because these are controlled modifications, we are able to isolate the effects of two closely-spaced minor merger events, the relative timing of which determines whether the MW-mass main galaxy quenches. This one-two punch works to:  1. fuel the supermassive black hole (SMBH) at its peak accretion rate; and 2. disrupt the cold, gaseous disk of the host galaxy. The end result is that feedback from the SMBH thoroughly and abruptly ends the star formation of the galaxy by $z\approx1$. We search for and find a similar quenching event in {\sc Romulus25}, a hydrodynamical $(25\,\mathrm{Mpc})^3$ volume simulation, demonstrating that the mechanism is common enough to occur even in a small sample of MW-mass quenched galaxies at $z=0$. 


\end{abstract}


\section{Introduction} 
\label{sec-intro}

Benchmark astronomical surveys, such as the Sloan Digital Sky Survey (SDSS) and the Cosmological Evolution Survey (COSMOS), have revealed how the bi-modality in galaxy properties evolves over redshift \citep[e.g.][]{Bell2003, Baldry2004, Strateva2001, Brammer2010, Ilbert2010, Muzzin2013}. Actively star-forming galaxies and non-star-forming, ``passive" galaxies occupy two distinct regions of parameter space in color magnitude diagrams and exhibit distinct morphologies \citep[e.g.][]{Wuyts2011,VanderWel2014} and stellar populations \citep[e.g.][]{McGee2011,Wetzel2012,LopezFernandez2018,Kauffman2003,Gallazzi2008}. Theoretical studies have been able to reproduce the bimodal galaxy distributions in SFR \citep{Feldmann2017}, morphology \citep{Snyder2015}, and color \citep{Kang2005,Nelson2018}; however, no theoretical consensus has yet emerged to explain the increase of quenched galaxies observed from $z \sim 1$ to present day \citep{Bell2004,Ilbert2013}.

The general decline in star formation rate towards $z \sim 0$ has been well-described by observational studies \citep[e.g.][]{Noeske2007,LopezFernandez2018}, and this process is almost certainly influenced by a decrease in cool gas supply in the local universe \citep[e.g.][]{Putman2012a}. However, there are many large-scale and small-scale processes that can impact the star formation properties of a galaxy.

\cite{Peng2010} describes two main quenching pathways: environmental \citep{Kauffmann2004,Baldry2006,Bahe2015} and mass \citep{Kauffmann2003b} quenching. Examples of such quenching processes include halo quenching or starvation \textemdash two types of mass-quenching\textemdash each cite a specific source driving their quenching. 

For example, halo quenching relies on the long cooling times of high-temperature ($\sim10^{6}$ K) halo gas \citep{Rees1977,Kauffmann1993,Somerville2015}. As IGM gas enters a high mass ($M_{\rm halo}$ $\gtrsim$ 10$^{11}$ M$_{\odot}$) galaxy through filaments, it shock heats to the virial temperature of the galaxy \citep{Keres2005,Dekel2006}, ultimately depriving galaxies of their star-forming fuel. However, within these massive halos, star formation suppression through this mode can be less efficient in high baryon fraction galaxies \citep{Benson2003} and cool gas may still permeate through shocked regions and accrete onto the galaxy \citep{Brooks2009,Dekel2009,Nelson2013}.

Similarly, the process which includes the physical removal and suppression of the gaseous fuel of a galaxy is called ``starvation'' and occurs in both low and high mass galaxies. In low mass galaxies with small gravitational potential wells, star formation feedback processes \textemdash such as stellar winds, radiation, and energy ejected via supernovae \textemdash are powerful enough to strip galaxies of some or all of their gas \citep{Larson1974,Dekel1986}. In more massive galaxies, AGN feedback is a likely culprit for ejecting the cool gas from a galaxy disk through powerful outflows \citep{Fabian2012,Cicone2014,Feldmann2015,Feldmann2016} and enriching the circumgalactic medium (CGM) with metals formed in the disk \citep{Suresh2017,Nelson2018,Sanchez2019}. In some cases, the AGN feedback energy can be strong enough to expel gas out of the CGM into the ICM \citep{Oppenheimer2020}. Additionally, the large scale cooling regulation which occurs in the CGM may drive gas back towards the galaxy, which can then further fuel the AGN, in a self-regulating galactic fountain \citep{Gaspari2013, Voit2017, Tremblay2018,Angles2017,Nelson2019,Chadayammuri2020}. Nevertheless, the AGN alone may not be capable of fully quenching a galaxy \citep{DiMatteo2005,Pontzen2017,Trussler2020}, and observations of highly star-forming galaxies can still show significant AGN activity \citep{Nandra2007,Simmons2012,Rosario2013,Mullaney2015, Bruce2016}.

In both of these cases, halo quenching and starvation, the main source of quenching comes from a specific physical driver\textemdash long cooling times and feedback processes\textemdash while observational evidence shows that these sources of quenching can be disrupted by other galactic properties such as AGN feedback in bright SF galaxies. Furthermore, additional studies find that the combination of halo quenching and the AGN activity driving starvation can work together to reduce star formation in some galaxies \citep{Bower2017,McAlpine2020}. In our study, we focus on the combination of physical processes which drive galaxy quenching through galaxy mergers. A third quenching process described by \cite{Peng2010}, merger quenching is mostly independent of mass and is typically associated with a major merger resulting in the cessation of star formation in a galaxy \citep{Toomre1972,Springel2005,Cox2006,Gabor2010}. However, we uniquely investigate this type of quenching through a series of minor merger interactions, rather than as the result of a single major merger.


We use a carefully constructed set of initial conditions to study merger-driven quenching within a controlled environment. Our study follows that of \cite{Pontzen2017} (hereafter, P17) which investigates quenching by black hole feedback and merger effects in tandem. P17 utilize the `genetic modification' technique \citep[GM,][]{Roth2016} to create a suite of cosmological simulations of Milky Way-mass (MW-mass) halos at $z = 2$ with assembly histories that have been modified in controlled ways. The environment and assembly history of each galaxy in the suite was nearly identical, except for a significant merger event with varying mass ratio (1:10, 1:5, and 2:3) occurring at $z < 2$. The resulting physical state of the main galaxy ranges from star forming to temporarily-quenched to permanently quenched due to the interplay between the major merger and SMBH feedback. In the permanently-quenched case (2:3), P17 show that the combined effort of the merger and the SMBH feedback work together to halt star formation: the merger disrupts the disk while the AGN feedback ejects and heats some, but not all, of the cold disk gas. It is the lack of an orderly disk that prevents further star formation despite some cool gas remaining in the galaxy. 

We follow the methods of P17 to investigate quenching in a new suite of genetically modified MW-mass galaxies at $z = 0$. However, we note that P17 examines the major mergers likely to occur at high-z for more massive halos, while we focus on MW-mass galaxies which have more quiet recent histories, like our own MW. In this study, we examine the influence of a more minor modification within these simulations and show that two minor mergers can lead to a unique form of quenching in MW-mass galaxies. 

This paper is organized as follows: Section \ref{sec-model} details our set of simulations and describes the genetic modification process. Section \ref{sec-results} reports our findings and results. In Section \ref{sec-conclude},  we summarize our results and discuss the broader implications of our findings.

\begin{figure}[]
\begin{center}
\includegraphics[scale=0.47]{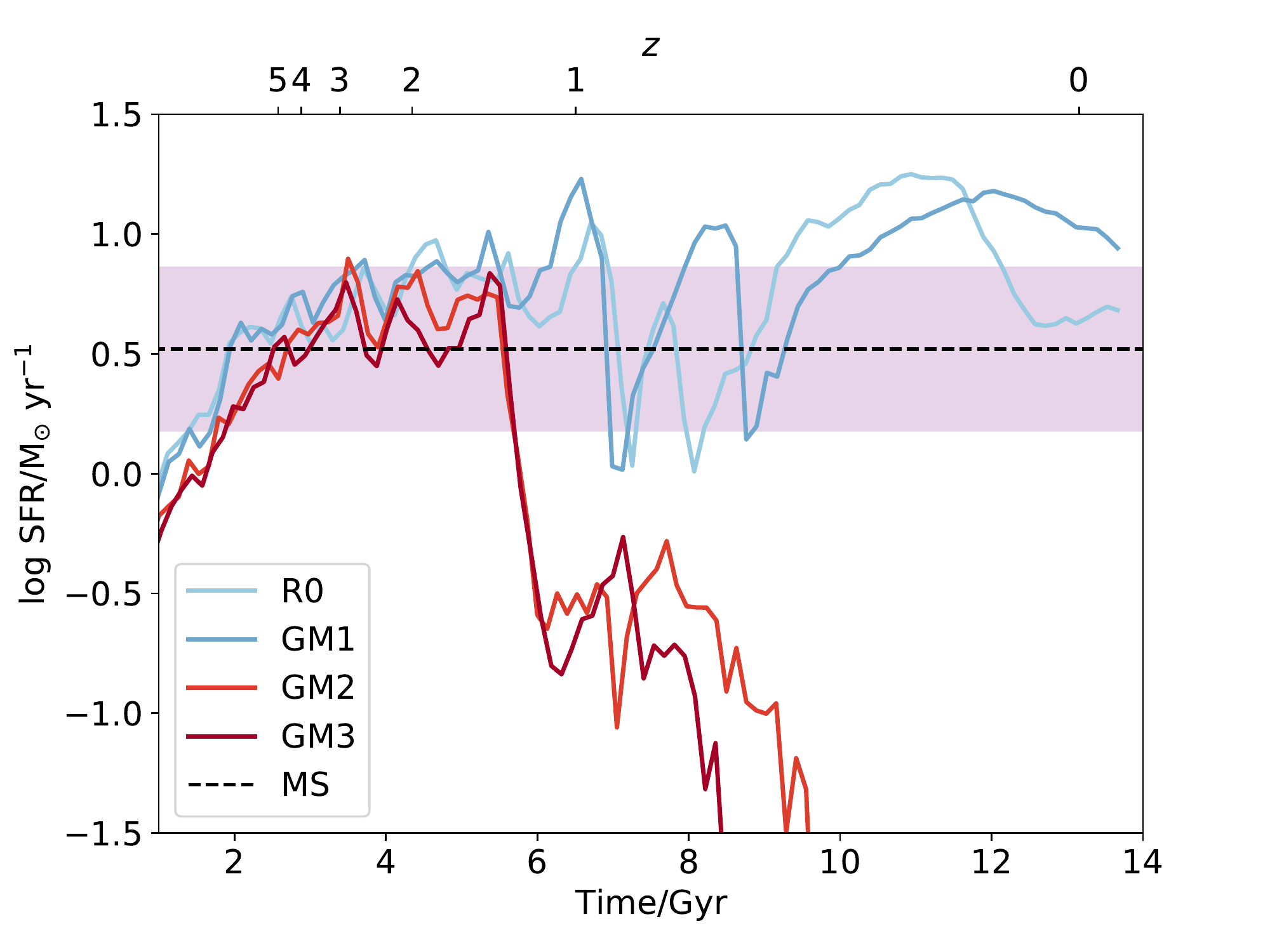}
\end{center}
\vspace{-3 mm}
\caption[]{Star formation histories of our 4 GM galaxies. The star forming galaxies, R0 and GM1, are shown in dark and light blue; the quenched galaxies, GM2 and GM3, are shown in dark and light red; and the main sequence star formation rate, for $M_{star} = 5 \times 10^{10} M_{\odot}$ at z = 0, is shown in purple \citep{Whitaker2012}. All 4 of our galaxies begin with very similar ICs which have been genetically modified to shrink the mass of a satellite which enters the main halo at $z = 1$. Despite their similar beginnings, two of the galaxies, R0 and GM1, remain star forming through their lives, while the others, GM2 and GM3, quench just after $z = 1$ and remain that way until $z = 0$ ($\sim 8$ Gyr).} 
\vspace{+3 mm}
\label{figure:sfh}
\end{figure}




\section{Simulation Parameters}
\label{sec-model}

To create our sample of galaxies, we used the modern SPH code, Charm N-body GrAvity solver \citep[ChaNGa][]{Menon2015}. ChaNGa inherents the same physical models as Gasoline \citep{Wadsley2004,Wadsley2017} and includes the following physical prescriptions: cosmic UV background \citep{Haardt2012}, star formation \citep[using an IMF given by][]{Kroupa2001a}, blastwave supernova feedback \citep[][for more details]{Ostriker1988,Stinson2012} including both SNIa and SNII \citep{Thielemann1986,Woosley1995}. SNII feedback imparts $10^{51}$ ergs of thermal energy per supernova onto surrounding gas particles. Low temperature metal line cooling \citep{Wadsley2008,Stinson2006a,Shen2012} is included to allow gas below $10^4 K$ to cool proportionally to the metals in the gas. Gas above this threshold cools only via H/He, Bremsstrahlung, and inverse Compton. No high temperature metal cooling is included due to the resolution of our simulations which does not resolve individual star forming regions \citep[See][for more detailed discussion]{Tremmel2019}.



Our simulations use an improved set of black hole (BH) prescriptions including formation, accretion, and dynamical friction \citep{Tremmel2017}. SMBH seeds form from dense, extremely low metallicity gas particles which allows BHs to form early in low mass halos, as predicted by the majority of theoretical models. Sub-grid models for SMBH accretion and dynamical friction have been implemented, including realistic SMBH mergers and dynamical evolution. SMBH dynamics are accurately followed down to sub-kpc scales \citep{Tremmel2015}. In particular, the subgrid model for accretion takes into account angular momentum support from nearby gas particles. This model allows for more physical growth compared to strictly Bondi-Hoyle accretion and does not require additional assumptions or free parameters. Angular momentum support is included in the accretion equation:

\begin{equation}
\dot{M} \propto \frac{\pi (GM_{BH})^2 \rho c_s}{(v_{\theta}^2 + c_s^2)^2},
\end{equation}

where $\rho$ is the density of the surrounding gas, $c_s$ is the sound speed, and $v_{\theta}$ is the rotational velocity of the surrounding gas. The quantity $v_{\theta}$ is informed by the angular momentum support of this gas on the smallest, resolvable scale. Additionally, a density dependent boost factor is implemented to avoid underestimating SMBH accretion rates due to resolution affecting temperature and density calculations of nearby gas. Using the prescription of \cite{Booth2009}, the standard Bondi rate is scaled by a density dependent factor, $(n_{gas}/n_{*})^{\beta}$, where $n_{*}$ is the star formation density threshold and $\beta$ is a free parameter. Combined, the density dependent boost factor and inclusion of angular momentum support results in the full equation from \cite{Tremmel2017}:

\begin{equation}
\dot{M} = \alpha \times \begin{cases}
\frac{\pi(GM)^2 \rho}{(v_{\mathrm{bulk}}^2+c_s^2)^{3/2}} & \text{ if } v_{\mathrm{bulk}}>v_{\theta} \\ \\
\frac{\pi(GM)^2 \rho c_s}{(v_{\theta}^2+c_s^2)^{2}} & \text{ if }  v_{\mathrm{bulk}}<v_{\theta}
\end{cases};
\end{equation}
\begin{equation*}
 \alpha = \begin{cases}
\left ( \frac{n}{n_{th,*}} \right )^\beta & \text{if } n \geq n_{th,*}\\ \\
1 &  \text{if } n < n_{th, \star}
\end{cases}
\end{equation*}

\noindent where $v_{bulk}$ is the smallest relative velocity of the SMBH's 32 nearest gas particles. Thus, in cases where bulk motions dominate over rotational motion, the formula reverts to Bondi-Hoyle.

Thermal SMBH feedback energy is determined by the accreted mass, $\dot{M}$, and imparted on the nearest 32 gas particles according to a kernel smoothing: 
\begin{equation}
E = \epsilon_{r} \epsilon_{f} \dot{M} c^2 dt, 
\end{equation}
where $\epsilon_{r}$ = 0.1 and $\epsilon_{f}$ = 0.02 are the radiative and feedback efficiency, respectively. Accretion is assumed to be constant over one black hole timestep, $dt$. Cooling is shut off immediately after AGN feedback events for a short ($\sim 10^{4-5}$ years) time. These choices were calibrated against dozens of zoom-in simulations to broadly reproduce observed galaxy and SMBH scaling relations. Furthermore, this SMBH feedback prescription is shown to produce large scale outflows \citep{Pontzen2017,Tremmel2019}. For more details on the SMBH prescriptions, see \cite{Tremmel2017}. 

Our simulations were each run with the same $\Lambda$CDM cosmology, $\Omega_{m}$ = 0.3086, $\Omega_{\Lambda}$ = 0.6914, $h$ = 0.67, $\sigma_{8}$ = 0.77 \citep{Planck2015}, and have a Plummer-equivalent softening length of 250 pc (a spline kernel of 350 pc is used). Initial conditions were generated using genetIC \citep{2020ascl.soft06020S}.

\begin{figure}[t]
\vspace{-3mm}
\begin{center}
\includegraphics[scale=0.45]{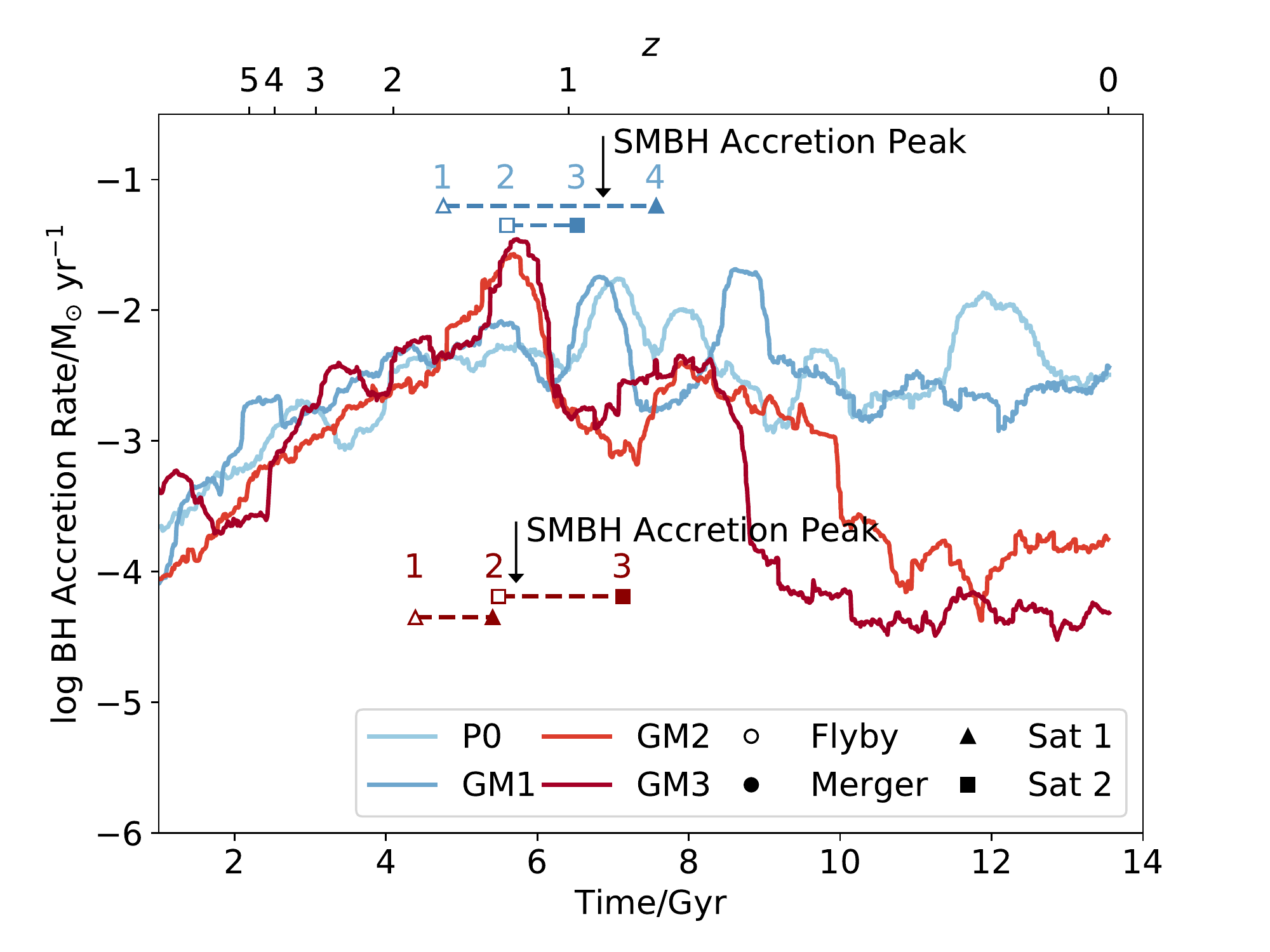}
\end{center}
\vspace{-3 mm}
\caption[]{SMBH accretion rates for our four GM galaxies across time averaged using a rolling mean of 450 Myrs. The peak of accretion and feedback energy occurs earlier in both quenched galaxies, within $\sim 100$ Myr of the double satellite interaction ($z \sim 1$). 
The timing of the satellite interactions are indicated in blue for the star forming galaxies, and red for the quenched cases. Triangles connected with a dashed line indicate the interaction period of the first satellite, from initial flyby (open marker) to time of its merger with the main halo (solid marker). For the second satellite interaction, open and closed squares indicate the initial flyby and time of merger, respectively. See Table \ref{table:mergerdata} for flyby and merger times.}
\label{figure:BHaccr_mass}
\end{figure}

\subsection{Halo and Merger Identification}
Individual halos are selected using the post-processing tool {\sc Amiga Halo Finder}, which selects halos using an overdensity criteria and grid based system which iteratively removes particles that are gravitationally unbound from prospective halos \citep{Knebe2001,Knollmann2009,Gill2004}. Virial mass, $M_{vir}$, and virial radii, $R_{vir}$, are determined using a spherical top-hat collapse technique. Halos are traced backwards through time from $z = 0$ to previous snapshots, following the halo from the previous snapshot which contains the majority of the same particles using the analysis tools pynbody \citep{pynbody} and TANGOS \citep{TANGOS}.

Merger ratios are defined at infall, during the snapshot just before the center of the satellite halo has first passed into the virial radius of the main halo (Table \ref{table:mergerdata}), and are calculated as $q = M_{vir,halo}/M_{vir,satellite}$. Larger infall ratios indicate mergers with smaller satellite galaxies.

These following zoom-in simulations were first described in \cite{Sanchez2019}, which compared the \ion{O}{6} column densities within these galaxies to observations from the COS-Halos Survey. This previous study examined the effects of star formation history and SMBH feedback on the circumgalactic medium. They found that while differences in the star formation histories of the galaxies didn't result in significant variations in the amount of \ion{O}{6} in the CGM of these galaxies, SMBH feedback was a significant driver of the metals into the CGM.

\subsection{The Genetic Modifications}
We selected a Milky Way-mass ``Organic'' galaxy, henceforth R0, ($M_{halo}$ = 9.9 $\times$ 10$^{11}$ M$_{\odot}$) from an initial, dark-matter  only cosmological volume which had uniform resolution and was 50 Mpc on a side. R0 was selected for the Large Magellanic Cloud-mass (LMC-mass, $M_{sat}$ = 2 $\times$ 10$^{10}$ M$_{\odot}$) satellite galaxy which was contained within its virial radius at $z = 0$, and was otherwise isolated (\textgreater 2 Mpc) from other MW-mass galaxies. Once selected, we define a Lagrangian region associated with this halo to create the ''zoom-in`` simulation of our R0 galaxy using the technique of \cite{Katz1993}. This zoom-in R0 includes baryons and their related physics while only re-simulating a few virial radii from the main halo at the highest resolution ($M_{gas}$ = 2.1 $\times$ 10$^5$ M$_{\odot}$, $M_{DM}$ = 1.4 $\times$ 10$^5$ M$_{\odot}$) while large scale structure at farther distances are simulated only in DM at a much coarser resolution.\footnote{Correction: \cite{Sanchez2019} states that the DM mass is 3.4 $\times$ 10$^5$ M$_{\odot}$ which is the DM mass resolution for ROMULUS25 not the GM galaxies.}

To create the subsequent ``genetically modified'' galaxies, we used the method of \cite{Roth2016} to modify the initial conditions of R0 by decreasing the mean over-density associated with the particles in the LMC-mass satellite which was present in R0 at $z = 0$. With this method, we created three GM galaxies (GM1, GM2, and GM3), each modified to result in a subsequently smaller satellite mass \citep[see Table 1; ][]{Sanchez2019}. The benefit of this method is that it allows us to fix the large scale structure and the final mass of the main halo ($M_{vir}\sim 10^{12} M_{\odot}$) while varying specific aspects of the halos assembly history. The simulations resulted in a set of four galaxies which, despite controlling for large scale environment and only slightly modifying the assembly of the halo, have varying baryonic evolution. Two of these galaxies, R0 and GM1, are star forming, disk galaxies, similar to the Milky Way, while two of these galaxies, GM2 and GM3, unexpectedly become quenched at $z \sim 1$ (Figure \ref{figure:sfh}).

Simulation snapshots of particle data had varying cadences with medians of 700 Myr and 200 Myrs for R0 and GM1, respectively, and 400 Myrs for both GM2 and GM3. Additional static images were created on the fly during each simulation with a cadence of 3 Myrs.

While first introduced in \cite{Sanchez2019}, two of these zoom-in simulations (GM2 and GM3 ) were additionally discussed in \cite{Cruz2020}, which examined the effect of self-interacting dark matter models on SMBH growth histories. Though the effects of varying assembly and star formation have been explored in these papers, no thorough treatment describing the quenching in these galaxies has yet been put forth. The purpose of the present paper is to explore the physical processes driving quenching in these galaxies.

\subsection{The {\sc Romulus25} Cosmological Volume}
{\sc Romulus25} \citep[][hereafter R25]{Tremmel2017} is a 25 Mpc cosmological volume that includes galaxies between halo masses of 10$^9$ \textemdash 10$^{13}$ M$_{\odot}$. The galaxies in R25 have been shown to lie along the $M_{BH}$-$M_{*}$,  stellar mass-halo mass, and $M_{BH}$-$\sigma$ relations \citep{Ricarte2019}, and are consistent with observations of of star formation and SMBH accretion histories at high redshift \citep{Tremmel2017}. Furthermore, \cite{Tremmel2017} shows that SMBH physics plays a necessary role in reproducing MW-mass galaxy evolution and quenching in high mass galaxies. R25 has a mass resolution of $M_{gas}$ = 2.1 $\times$ 10$^5$ M$_{\odot}$ and $M_{DM}$ = 3.4 $\times$ 10$^5$ M$_{\odot}$ for gas and DM particles, respectively.

For our study, we examine a set of 26 MW-mass galaxies which have final halo masses, M$_{halo}$, between 5 $\times$ 10$^{11}$ and 2 $\times$ 10$^{12}$ M$_{\odot}$ and
which are not satellites of a more massive halo at $z = 0$. Our M$_{halo}$ measurements use the corrections of \cite{Munshi2013}.


\section{Results}
\label{sec-results}


\begin{table*}[] 
\caption{Timing of the Minor Merger Scenarios} 
\centering 
\begin{tabular}{c c c c c c c c} 
\hline\hline 
Sim & Sat 1  & Sat 1  & Sat 1  & Sat 2  & Sat 2 & Sat 2  & SMBH  \\
& Infall & Flyby  & Merger & Infall & Flyby & Merger & Accretion \\
& Ratio  &        &        & Ratio  &       &        & Peak \\
& q       & Gyr    & Gyr   &  q      & Gyr   & Gyr    & Gyr \\ [0.5ex]
\hline 
R0  & 5.4 & 4.76 & 7.57 & 13.6 & 5.60 & 6.53 & 6.95  \\ 
GM1 & 7.3 & 4.69 & 7.39 & 14.6 & 5.40 & 6.35 & 6.74  \\
GM2 & 8.2 & 4.43 & 5.42 & 18.9 & 5.44 & 6.80 & 5.59  \\ 
GM3 & 9.5 & 4.39 & 5.41 & 17.9 & 5.49 & 7.13 & 5.84  \\ [1ex] 
\end{tabular} \\
Details about satellite interactions in our four GM galaxies, including infall merger ratios, flyby times, merger times, as well as the time of the peak accretion rate of the SMBH. Infall merger ratios, q, are defined as $M_{vir,halo}/M_{vir,satellite}$ at the simulation output before the satellite enters the main halo. \\ Flyby and merger times were determined by visual examination of ppm image files created on the fly during simulation with a cadence $\sim$ 3 Myr.
\label{table:mergerdata} 
\end{table*}

\begin{table*}[] 
\caption{Properties of Zoom-In Galaxies Prior to Minor Merger Interactions\\}
\centering 
\begin{tabular}{c c c c c c c c} 
\hline\hline 
Sim & Total Halo Mass  & Total Gas Mass & Total Stellar Mass & Cold Gas Mass & Dense Gas & R$_{vir}$ & T$_{vir}$ \\
 & (M$_{\odot}$) & (M$_{\odot}$) & (M$_{\odot}$) & (M$_{\odot}$) & (M$_{\odot}$) & (kpc) & (K) \\ [0.5ex] 
\hline 
R0 ($z \sim 1.18$) & 5.7 $\times$ 10$^{11}$ & 7.1 $\times$ 10$^{10}$ & 1.3 $\times$ 10$^{10}$ & 4.1 $\times$ 10$^{9}$ & 7.0 $\times$ 10$^{9}$ & 122.8 & 8.3 $\times$ 10$^5$ \\ 
GM1 ($z \sim 1.32$) & 4.9 $\times$ 10$^{11}$ & 6.0 $\times$ 10$^{10}$ & 1.2 $\times$ 10$^{10}$ & 3.7 $\times$ 10$^{9}$ & 7.0 $\times$ 10$^{9}$ & 110.0 & 8.4 $\times$ 10$^5$ \\
GM2 ($z \sim 1.32$) & 5.1 $\times$ 10$^{11}$ & 6.4 $\times$ 10$^{10}$ & 9.1 $\times$ 10$^{9}$ & 5.0 $\times$ 10$^{9}$ & 4.6 $\times$ 10$^{9}$ & 111.4 & 8.6 $\times$ 10$^5$ \\ 
GM3 ($z \sim 1.32$) & 5.0 $\times$ 10$^{11}$ & 6.0 $\times$ 10$^{10}$ & 7.5 $\times$ 10$^{9}$ & 5.0 $\times$ 10$^{9}$ & 3.3 $\times$ 10$^{9}$ & 110.5 & 8.4 $\times$ 10$^5$ \\ [1ex] 
\end{tabular} \\
Details about the simulations prior to the beginning of the minor merger interactions at $z \sim 1$, including total virial halo mass, total gas mass, total stellar mass, cold gas mass and dense gas mass, all in the main halo. Virial radius and virial temperature of the halo are also included. The properties of R0 are shown at $z \sim 1.18$ due to the limited number of simulation outputs available for this simulation. The properties of the three other simulations are shown at $z \sim 1.32$.
\label{table:gxydata} 
\end{table*}

\subsection{Differences in Merger Timings}
\label{sec-results1}

Due to the constraints that maintain the final mass of the main halo while changing the LMC satellite mass, the genetic modification technique affects the timing of accretion throughout the evolution of the galaxy. In GM2 and GM3, our two quenched cases and those with the smallest satellite masses, the accretion of satellites onto the main galaxy must occur faster and therefore earlier to maintain the final mass of the main halo. Consistently, SMBH accretion also peaks earlier ($z \sim$ 1.18), nearly 1 Gyr before the peak of SMBH accretion in the two star forming cases, R0 and GM1 (Figure \ref{figure:BHaccr_mass}, discussed in detail below). The differences in timing and order of the minor mergers which occur are key to understanding the effect of the quenching in these two galaxies. 

We note that the variations between the two star forming galaxies themselves are minimal. Similarly, the two quenched cases have timing and sequence that are closely similar (Table \ref{table:mergerdata}). For that reason, we will generalize to two cases: the star forming case and the quenched case. 

\begin{figure*}[]
\vspace{-3 mm}
\begin{center}
\includegraphics[scale=0.13]{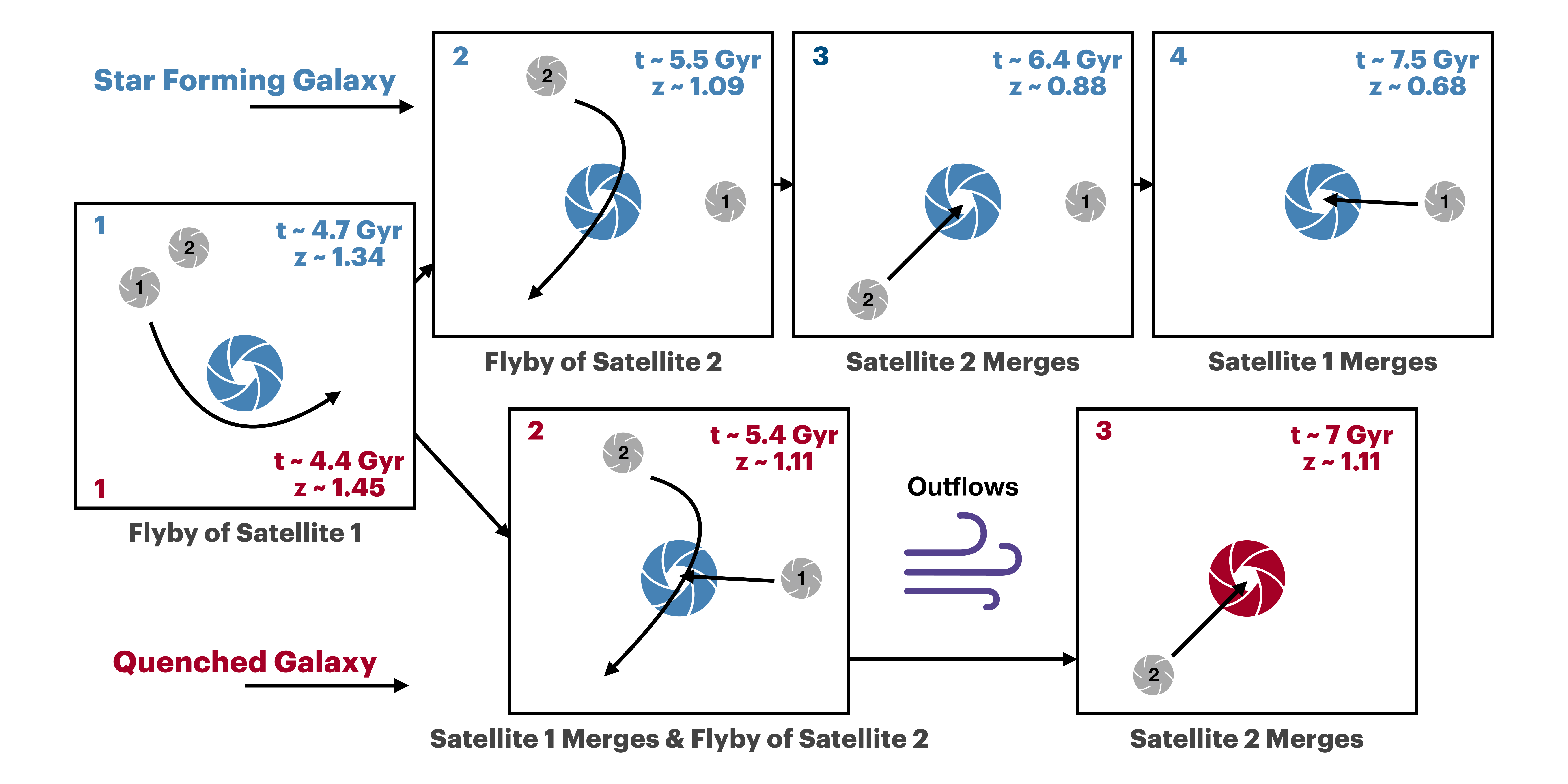}
\end{center}
\vspace{-3 mm}
\caption[]{Diagram detailing the order of the satellite merger scenario in the star forming (\textit{Upper}) and quenched (\textit{Lower}) cases. In the star forming case, the satellite interactions occur in this order: flyby of satellite 1, flyby of satellite 2, then the merger of satellite 2, and finally satellite 1 merges last. In the quenched case, the order of these interactions is different. Satellite 1 still interacts with a flyby first, however it then merges with the main halo within a Gyr. Shortly after (\textless 100 Myr), the flyby of satellite 2 occurs. Additionally, in the quenched case, the time when satellite 1 merges and satellite 2 does its flyby is shortly followed by the peak of SMBH accretion in these galaxies (a few 100 Myrs).}
\label{figure:mergerdiagram}
\end{figure*}

Figure \ref{figure:mergerdiagram} illustrates the differences in the order and timing of the minor mergers in the star forming (\textit{Upper Panels}) and quenched cases (\textit{Lower Panels}). In the star forming case, (1) satellite 1 and satellite 2 are both infalling towards the galaxy of the main halo by $t \sim 4.7$ Gyr, when satellite 1 does a flyby of the main galaxy. (2) Satellite 2 then does a flyby nearly a Gyr later at $t \sim 5.5$ Gyr. (3) Satellite 2 merges another Gyr after that at $t \sim 6.4$ Gyr. (4) Finally, satellite 1 merges last at nearly $t \sim 7$ Gyr.

In the quenched case, the order and timing of these same interactions are markedly different. (1) Satellite 1 does its flyby nearly half a Gyr earlier ($t \sim 4.4$ Gyr) than in the star forming case, consistent with the earlier, faster accretion expected from the galaxies with the most significantly shrunken satellite mass. (2) A Gyr after the flyby of satellite 1, satellite 1 merges with the main halo at $t \sim 5.4$ Gyr and the flyby of satellite 2 quickly follows, occurring within the next 100 Myr. In the quenched case, (3) satellite 2 is the last to merge, a little more than 1.5 Gyr after the double interaction preceding it. The specific order and timing between these interactions are what set the stage for the stark result of quenching in this galaxy rather than continued star formation.

\begin{figure*}[]
\vspace{-2mm}
\begin{center}
\includegraphics[scale=0.7]{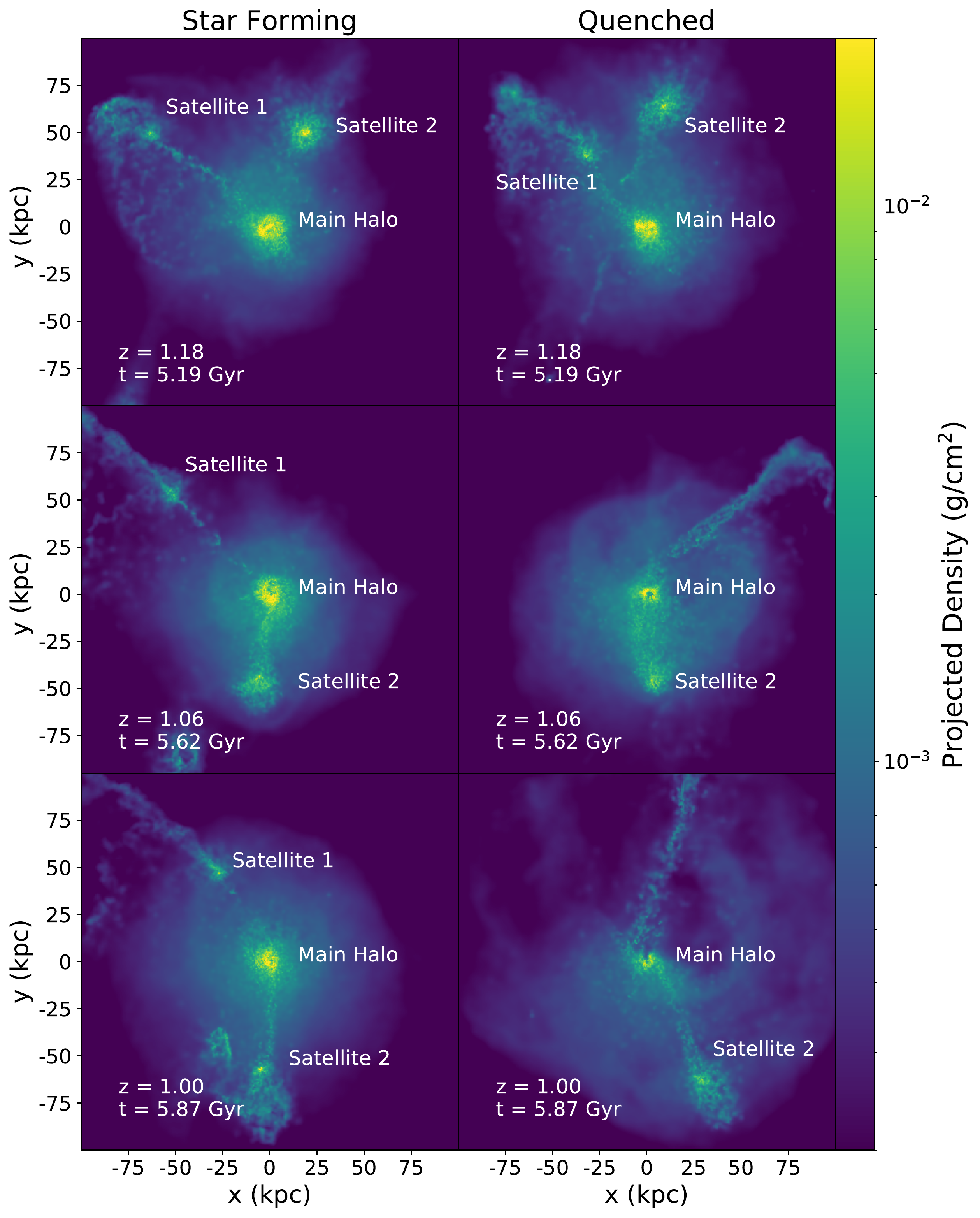}
\end{center}
\caption[]{Gas density maps of the star forming galaxy, GM1, and quenched galaxy, GM2, around the times of the minor satellite interactions. \textit{Left:} In GM1, satellite 1 and 2 are both infalling at z = 1.18 (\textit{Top Panel}). About a half a Gyr later (\textit{Middel Panel}), satellite 2 has completed its flyby and satellite 1 is still infalling towards the main halo. By $z = 1$, both satellites are making their way toward the main halo where they will finally merge around t $\sim$ 7.5 and t $\sim$ 6.5 for satellite 1 and 2, respectively. \textit{Right:} In GM2, one of our quenched galaxies, the order and timing of these interactions have some key differences. At $z = 1.18$ (\textit{Top Panel}), like in the star forming GM1, the main halo of GM2 has experienced the flyby of satellite 1, while satellite 2 is still in its initial infall. However by $z = 1.06$ (\textit{Middel Panel}), satellite 1 has fully merged with the main halo and satellite 2 has completed its flyby, in contrast to the star forming case which still shows both satellites. Finally at $z = 0$ (\textit{Bottom Panel}), satellite 2 is infalling back towards the main halo and will merge with it in about 1.5 Gyr.}
\label{figure:mergerrhomaps}
\end{figure*}

Figure \ref{figure:mergerrhomaps} includes a series of gas density maps spanning the time of these interactions.  GM1, our star forming case, is shown on the left, while GM2, our quenched case is on the right. At t $\sim$ 5.2 Gyr in the star forming case (\textit{Upper Left}), satellite 1 has completed its flyby of the main galaxy and is still moving away from it, while satellite 2 is infalling. At this time, our quenched GM2 (\textit{Upper Right}) has experienced the same interaction. However, by t $\sim$ 5.6 (\textit{Middle Left Panel}), satellite 2 in GM1 has completed its flyby of the main galaxy and satellite 1 is falling back towards it returning from its initial flyby. In contrast (\textit{Middle Right Panel}), satellite 1 in GM2 has fully merged with the main halo by this time, with satellite 2 having completed its flyby as well. Finally, at t $\sim$ 5.9 (\textit{Bottom Left Panel}), both GM1 satellites are now infalling back towards the main galaxy. In GM2, (\textit{Bottom Right Panel}), satellite 2 alone is infalling and will complete its merger in about another Gyr.

\begin{figure*}[]
\vspace{0mm}
\begin{center}
\includegraphics[scale=0.35]{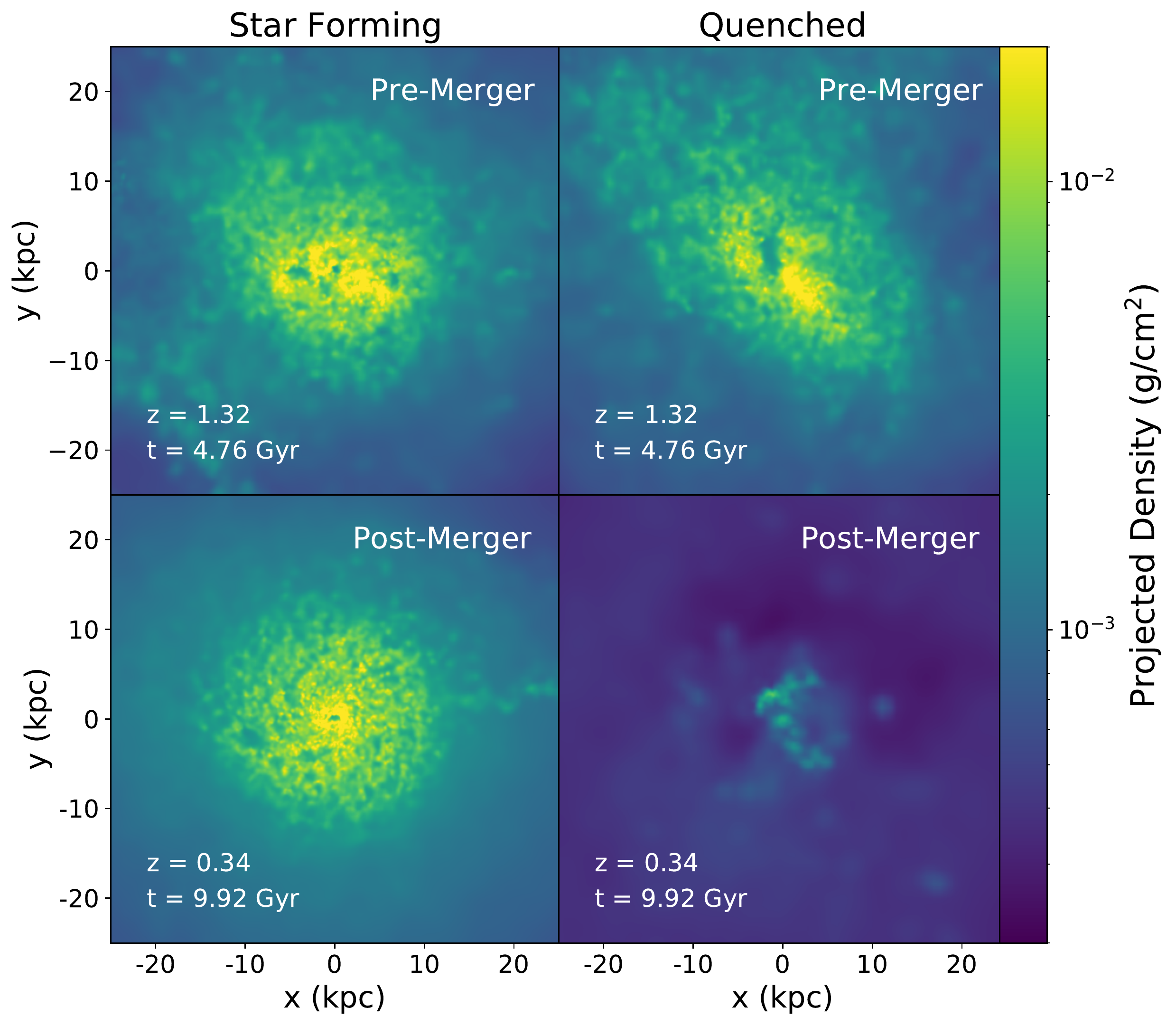}
\hspace{-3 mm}
\includegraphics[scale=0.35]{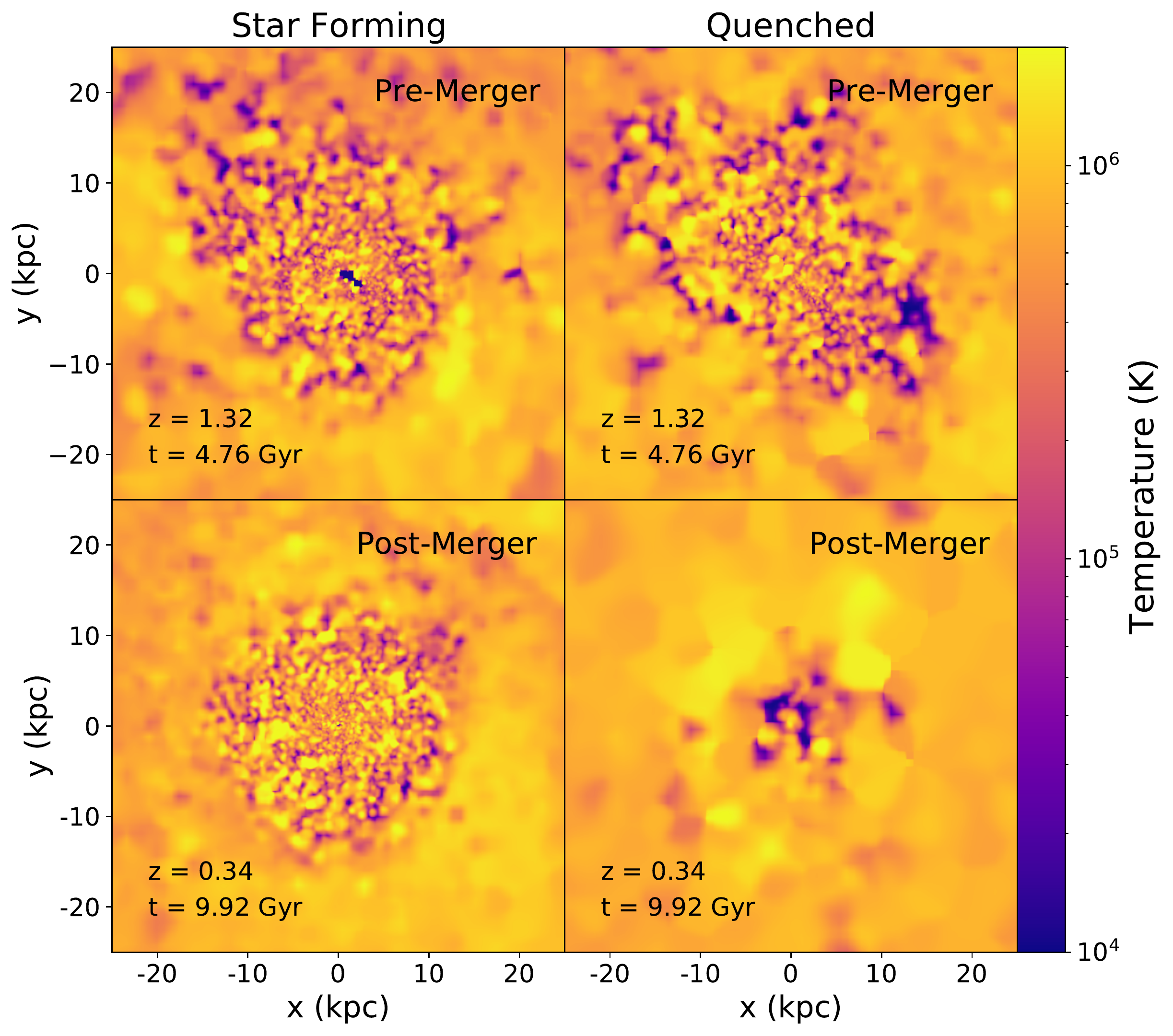}
\end{center}
\caption[]{Gas density (\textit{Left}) and temperature maps (\textit{Right}) of the star forming galaxy, GM1, and quenched galaxy, GM2. The upper panels show both galaxies long before the minor merger interactions which quenched GM2 while both galaxies experience a time of disk stability. The lower panels of each galaxy show them at a time long after the interaction has impacted the galaxies, showing the stable disk that GM1 has maintained through the series of interaction at $z = 1$ and the complete lack of disk and cold gas in GM2.}
\label{figure:rhotempmaps}
\end{figure*}

\begin{figure}[ht!]
\begin{center}
\hspace*{-5mm} 
\includegraphics[scale=0.48]{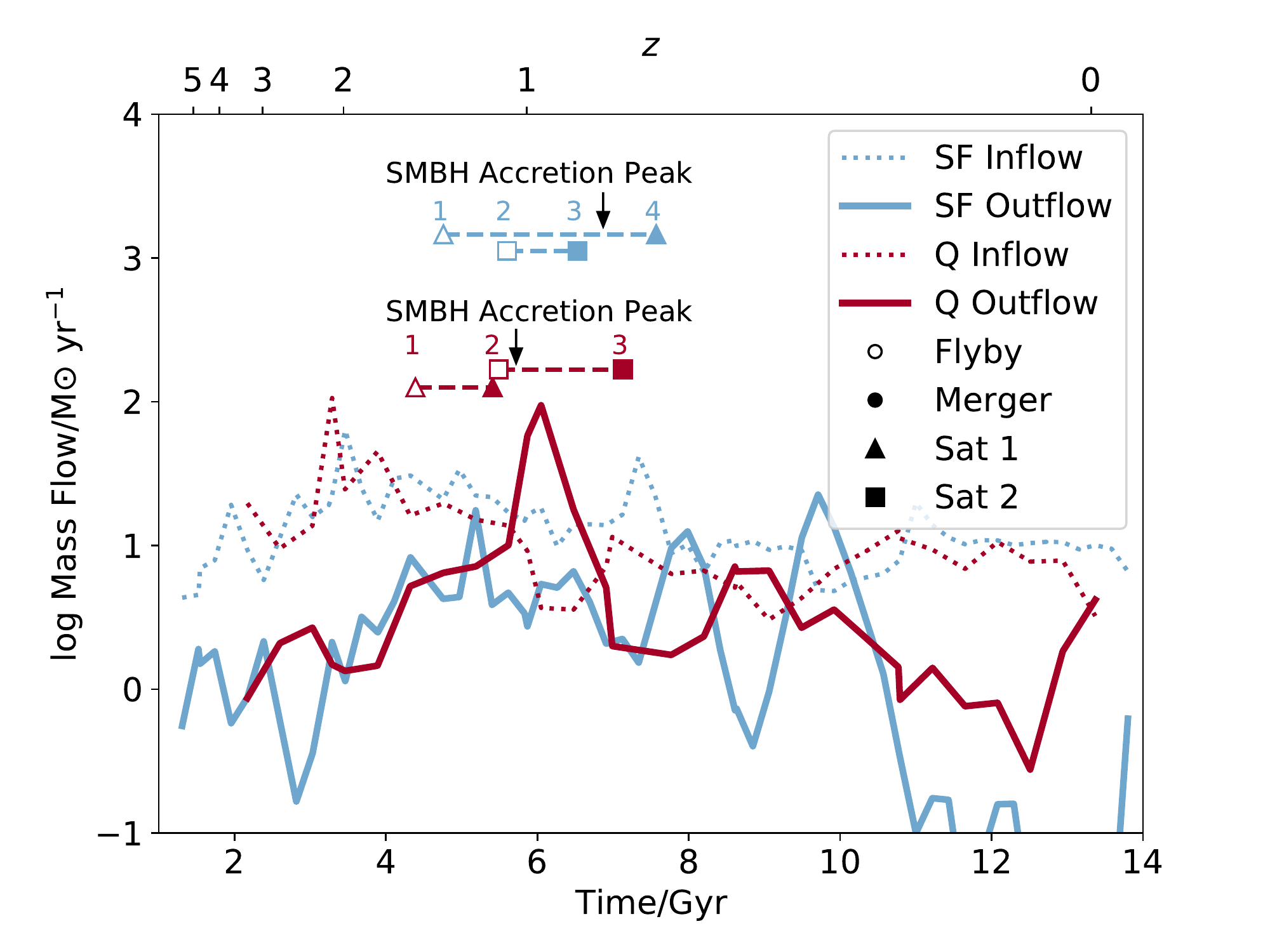}
\end{center}
\caption[]{The mass flow as a function of time in GM1, the star forming case, and GM2, the quenched case. Blue lines denote GM1, one of our star forming galaxies, and red lines denote GM2, one of the quenched cases. Dotted lines indicate inflow at the virial radius, while solid lines indicate outflow at the virial radius. The minor satellite flybys and mergers are indicated as in Figure \ref{figure:BHaccr_mass}. A significant outflow occurs at $z \sim 1$, directly following the minor satellite interaction (when satellite 1 merges and satellite 2 follows with a flyby) and the peak in SMBH accretion.}
\label{figure:massflow}
\end{figure}

\begin{figure}[h]
\vspace{-0mm}
\begin{center}
\includegraphics[scale=0.45]{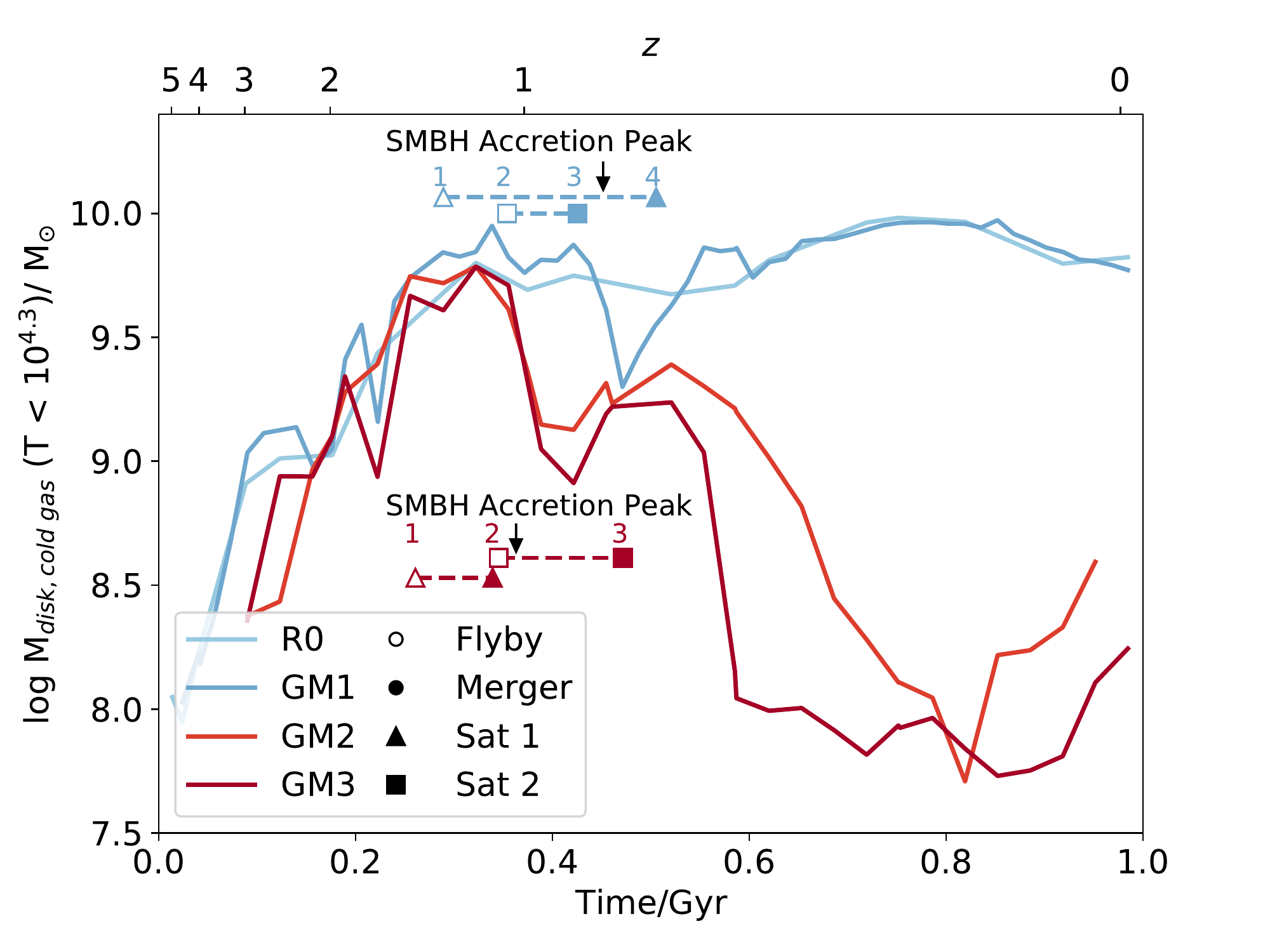}
\end{center}
\vspace{-10 mm}
\begin{center}
\includegraphics[scale=0.45]{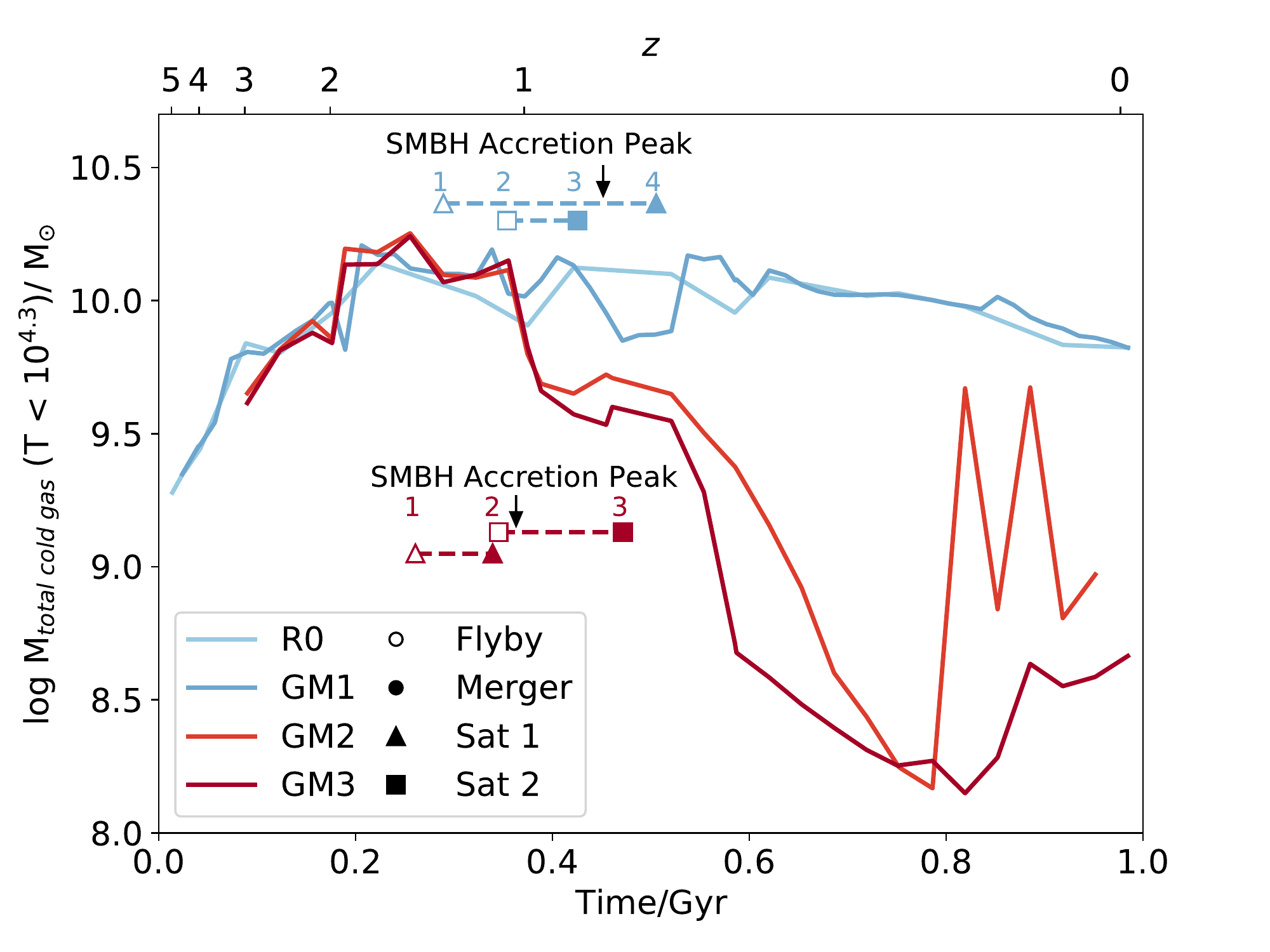}
\end{center}
\caption[]{Cold disk gas ($T <  2 \times 10^{4}$ K, $R < 0.1 R_{vir}$) and cold gas mass ($T <  2 \times 10^{4}$ K, $R < R_{vir}$) in our GM galaxies. \textit{Upper:} Prior to $\sim$ 6 Gyr, the amount of cold gas in the disk of the star forming and quenched galaxies is not significantly different. \textit{Lower:} Similarly, we see consistent amounts of total cold gas mass in all 4 of the halos prior to this time. However, in both figures, once the minor satellite interactions occur (red filled triangle and open red square) and the SMBH accretion rate peaks in GM2 and GM3 (lower down-turned black arrow), the majority of this cold gas is removed in a large outflow from the disk (Figure \ref{figure:massflow}). Line colors, styles, and marker styles as in (Figure \ref{figure:BHaccr_mass}).} 
\label{figure:diskcoldgasprof}
\end{figure}

\begin{figure*}[]
\vspace{-2mm}
\begin{center}
\includegraphics[scale=0.44]{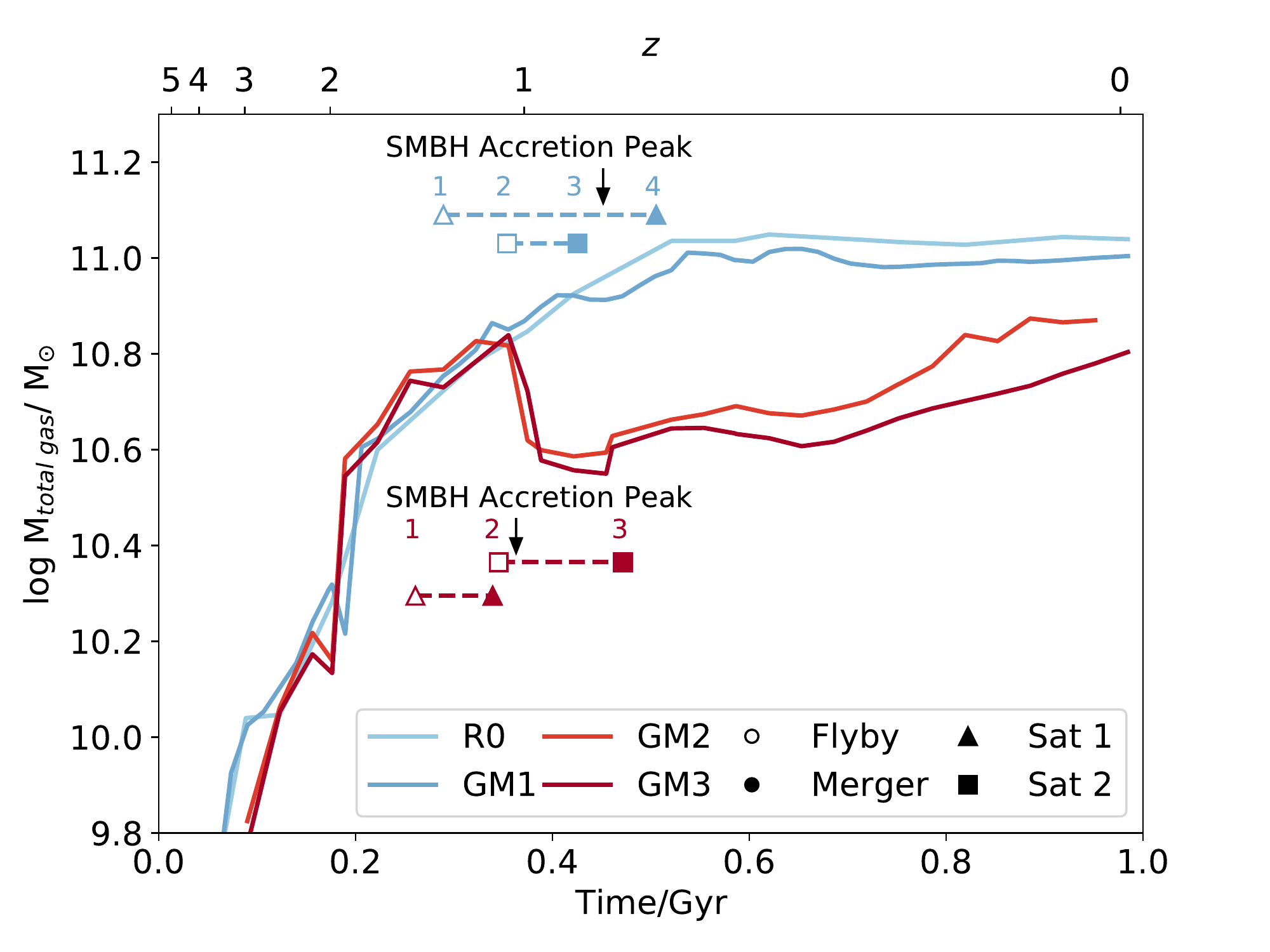}
\includegraphics[scale=0.44]{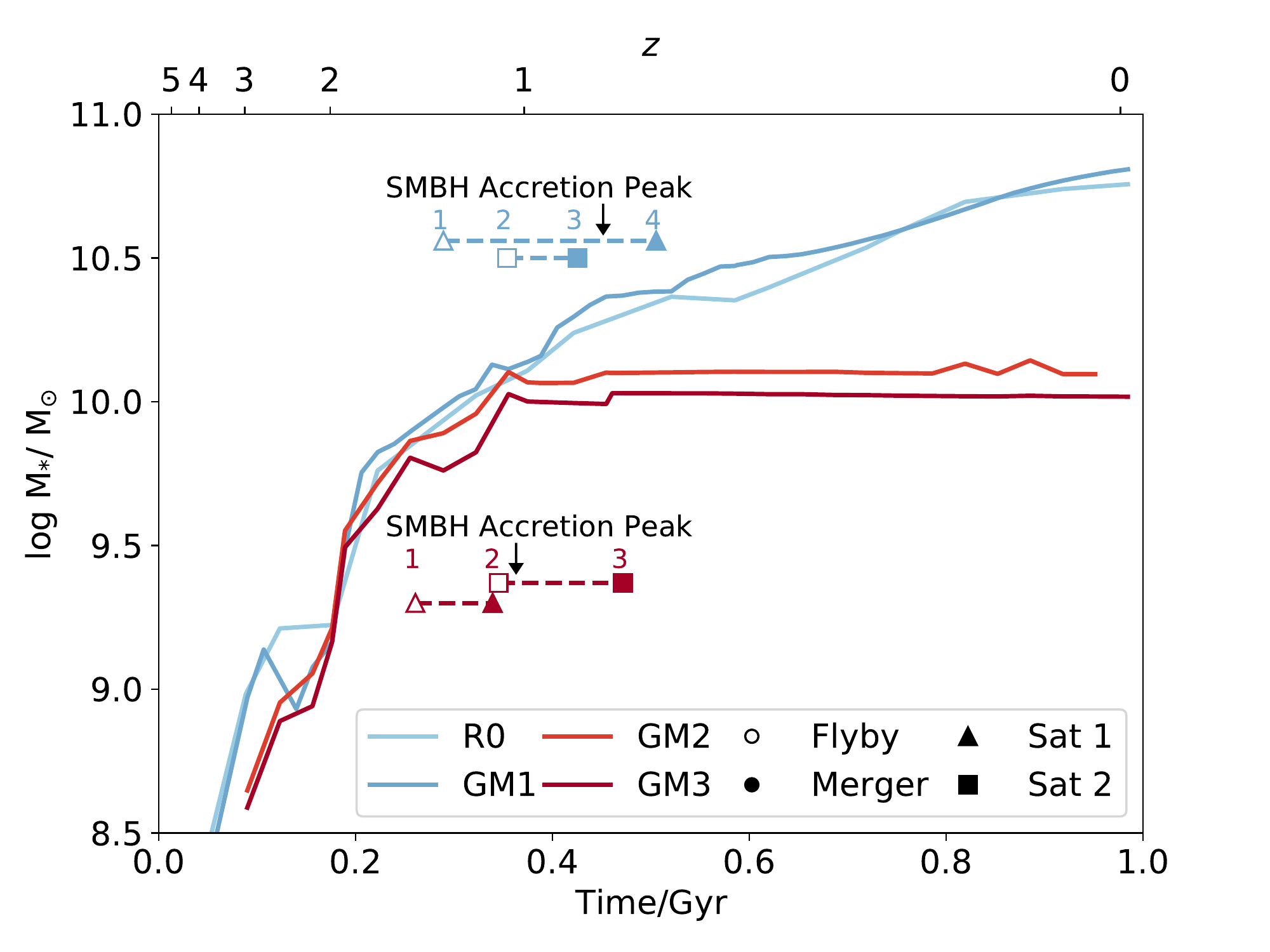}
\end{center}
\caption[]{The total gas and total stellar mass for the 4 GM galaxies. Prior to $\sim$ 6 Gyr, there is little variation in either the total gas mass (\textit{upper}) or stellar mass (\textit{lower}) the star forming or quenched galaxies. The key difference affecting the overall properties of the galaxies after $z \sim 1$ is the timing between the minor mergers. Line colors, styles, and marker styles as in (Figure \ref{figure:BHaccr_mass}). The timing between the mergers in the star forming case ($\sim$ 1 Gyr) is significantly longer than that of the quenched case ($\sim$ 100 Myr) in which the timing of the minor merger interaction coincides with the peak of SMBH accretion in the quenched case.}

\vspace{+3mm}
\label{figure:propertiesthrutime}
\end{figure*}

\subsection{Satellite Interactions and SMBH Feedback}
\label{sec-results2}

While we've shown that these interactions do not have a large effect on the star forming cases, R0 and GM1, both GM2 and GM3 experience significant outflows at $z \sim 1$, quenching the galaxy completely for the rest of the simulation (Figure \ref{figure:massflow}). Outflows and inflows are calculated by measuring the velocity of the gas passing through a shell at the virial radius with thickness of $0.1R_{vir}$. While the inflow rates of the galaxies generally follow a similar shape (dotted lines), a clear and significant difference is present in the outflows (solid lines). While there is no large outflow in the star forming case (blue solid line), in the quenched case there is a large outflow (red solid line) directly following the minor satellite interactions and SMBH accreiton peak at $t \sim 6$ Gyr. These outflows expel most of the gas from the disk, removing the fuel supply for further star formation (Figure \ref{figure:diskcoldgasprof}).

We investigate the galaxy properties during the time just before quenching in GM2 and GM3 to understand why they quench while the others do not. Specific characteristics of the merger do not appear to be drivers of the quenching (Table \ref{table:gxydata}). No significant differences arise between the primary halos with regards to the total virial mass, gas mass, or stellar mass at the time of the merger or leading up to it. There is also no significant difference between the amount of cold ($< 2 \times$ 10$^4$ K) gas in the disk or the entire halo (Figure \ref{figure:diskcoldgasprof}). We do not find significant differences between the properties of the star forming and quenched galaxies prior to $z = 1$ (Figure \ref{figure:propertiesthrutime}), instead determining that the main difference between these GM simulations is directly related to the timing of their satellite interactions.

Figure \ref{figure:BHaccr_mass} not only shows the accretion rate of the SMBH at the center of each GM main galaxy, but additionally the timing and sequence of the satellite interactions are marked. In the star forming cases (solid lines in blue), the mass of the SMBH continues to grow during the times of the mergers ($z \sim 1$) and the first peak of SMBH accretion occurs at $t \sim 6.8$ Gyr. This peak in accretion coincides closely with the merger of satellite 2, but without disrupting star formation. In comparison, the peak of SMBH accretion in the quenched cases occurs at $t \sim 5.8$ Gyrs, following both the merger of satellite 1 and flyby of satellite 2 within a few Myrs. This set of interactions is followed by a significant outflow at $t \sim 6$ Gyr, after which the galaxies remain quenched for the rest of their lives.

\begin{figure*}[ht]

\begin{center}
\includegraphics[scale=0.6]{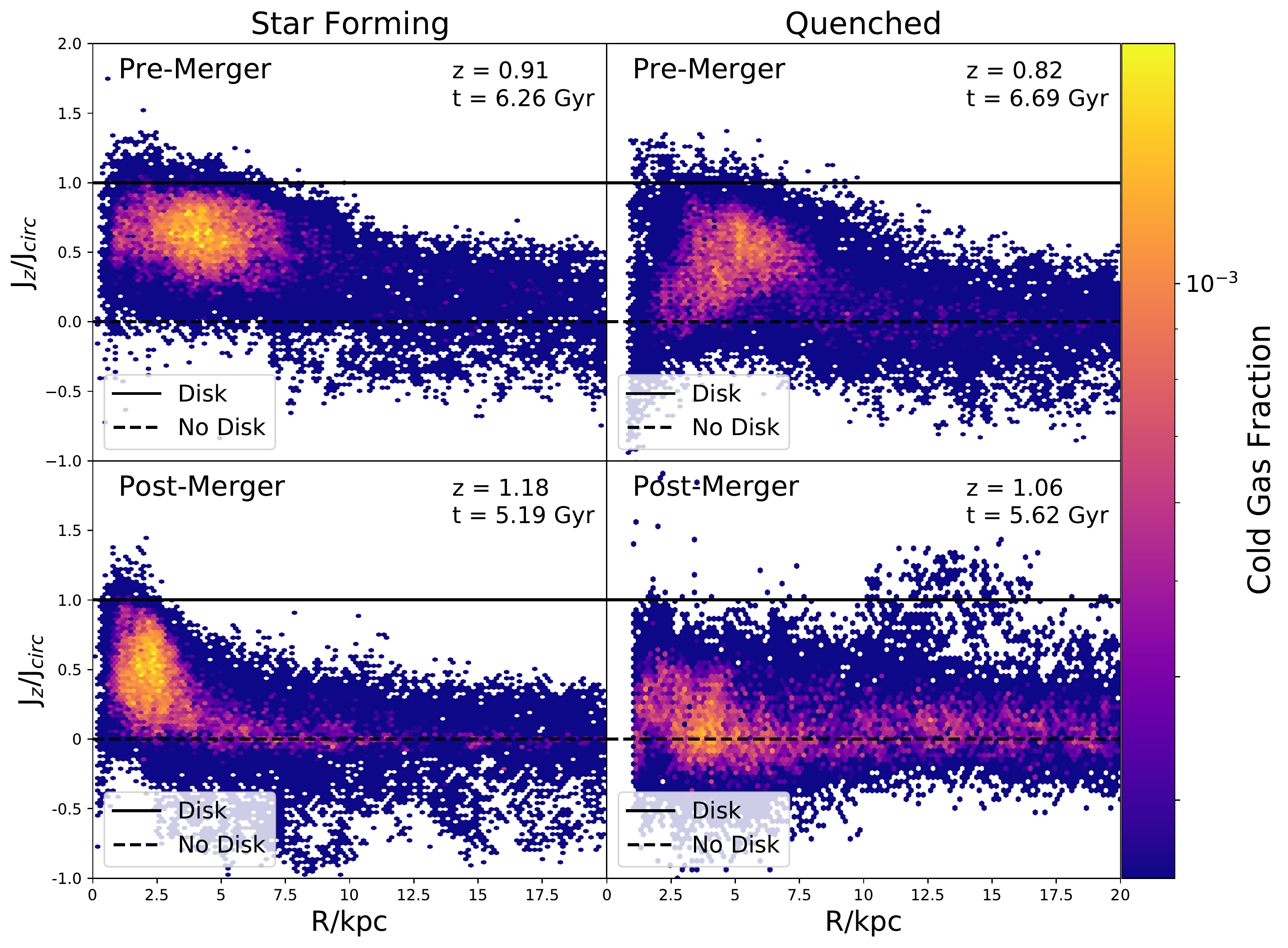}
\vspace{-5mm}
\end{center}
\caption[]{Plots of the circularity parameter $jz/jcirc$ of the cold gas ($T < $ 10$^5$ K) for the star forming MW-mass galaxy, GM1, and a quenched MW-mass galaxy, GM2, around the time of the satellite merger in each galaxy. Gas that has $jz/jcirc$ closer to 1 is rotationally supported (i.e. in a disk), while gas with $jz/jcirc$ closer to 0 is dispersion supported. \textit{(Top:)} Prior to the satellite mergers which result in GM2 quenching, both galaxies have fairly stable gaseous disk components. \textit{(Bottom:)} After the interaction and mergers occur, however, the star forming galaxy \textit{(Left)} retains a stable disk. You can also see compaction of the gas in the post-merger case \citep{Dekel2014}. Meanwhile the quenched galaxy \textit{(Right)} at the post-merger time lacks a stable cold gas disk.}
\label{figure:JzJcirc}
\vspace{+3 mm}
\end{figure*}

Given no significant differences in the physical characteristics of the galaxies, or their SMBHs, prior to the series of interactions that occur at $z \sim 1$, we then look to the dynamics of the disk to better understand how the differences in galaxy accretion history arise.
Figure \ref{figure:JzJcirc} shows the circularity parameter ($j_{z}/j_{circ}$, see \cite{Keller2015,Simons2019} and references therein) of cold gas ($T < 2 \times 10^4$ K) for the star forming, R0, and quenched, GM2. The galaxy disk is first oriented on the total angular momentum of the gas within 5 kpc of the galaxy center. Then for each cold gas particle within 20 kpc, the circularity parameter is calculated as ratio of its specific angular momentum component perpendicular to the disk ($j_{z}$) and the specific angular momentum for the theoretical circular orbit of that particle in its current potential  ($j_{circ}$). Values of $j_{z}/j_{circ}$ closer to 1 indicate gas that is rotationally supported in a disk, while gas with $j_{z}/j_{circ}$ \textless\ 0.5 is dispersion dominated. Gas prior to the merger (\textit{upper panels}) is stable and mostly rotationally supported in both galaxies. Difference arise after the mergers occur (\textit{lower panels}). We see that GM1 (\textit{bottom left panel}), our star forming case,  retains a stable disk which has become compacted after the merger \citep{Dekel2014}. While GM2 (\textit{bottom right panel}), our quenched case, has cold gas which is no longer rotationally supported in a disk ($j_{z}/j_{circ}$ values closer to 0). This difference is a key component in our result.

\subsection{The Quenching Combination}
\label{sec-results3}

In both the GM cases where the galaxy quenches after $z = 1$, the difference in the satellite merger combination is present. Additionally, the earlier accretion of satellite 1 feeds the SMBH with its gas, resulting in an earlier peak of SMBH accretion than in the SF galaxies. We determine that the subsequent disruption of the disk in these quenched cases \textemdash through the minor merger interaction of the merger and subsequent flyby \textemdash allows the resulting SMBH feedback (from the peak of SMBH accretion) to eject a majority of the cold gas in the disk (Figure \ref{figure:diskcoldgasprof}). This one-two punch combination of minor satellite interactions, in tandem with the SMBH feedback, works to quench the galaxies until $z = 0$. In short, to quench these galaxies, the combination of fuel given to the SMBH by satellite 1 and the disruption of the main galaxy disk by both satellite interactions results in SMBH-driven outflows strong enough to quench the galaxy until $z = 0$.

Our result is broadly consistent with that of \cite{Pontzen2017}. Their results from a different set of genetically modified galaxies show that a disk instability resulting from an interaction is necessary for a galaxy to quench. Another requirement for quenching is the presence of a SMBH, and in particular they concluded that continuous bursts of feedback were necessary to maintain their quenched galaxies. \cite{Sanchez2019} further refined the latter requirement by examining R0 and the same GM galaxies explored in this paper, both with and without BHs. Unlike in P17 however, we find that a single burst of SMBH is enough to quench two of our galaxies for nearly 8 Gyr without further episodes of SMBH feedback. We attribute these varying results to the differences in redshift and galaxy mass in each study, ours exploring lower mass galaxies at low redshift.

\subsection{Quenching Galaxies in a Broader Context}

Each of our quenched galaxies has an interaction timescale of $\sim$ 100 Myrs between when satellite 1 merges and satellite 2 does its subsequent flyby. The timescale between the mergers of each satellite in the quenched cases are on the order of 1 Gyr (bottom panels in Figure  \ref{figure:mergerdiagram}). In contrast, the order and timing of the flybys and mergers are markedly different for the two cases in which star formation does not cease. While in these cases satellite 2 merges before satellite 1, these distinct events are separated by a similar 1 Gyr (top panels, Figure \ref{figure:mergerdiagram}). We use the timing constraints above to guide an analysis of quenching in the larger, cosmological simulation, {\sc Romulus25} \citep{Tremmel2017}. The purpose of this analysis is to understand the role of minor mergers in quenching MW-mass galaxies, which has thus far been unexplored.


To constrain how likely this type of event might be in the (z \textless 2) universe, we examine a population of MW-mass galaxies from the cosmological volume R25. While a larger DM-only volume may provide a more statistically significant measurement for how often these minor mergers occur in MW-mass galaxies overall, we choose instead to select our additional sample from R25. First, it provides a larger, uniform sample of isolated MW-mass galaxies with the same physics and resolution as the GM simulations. Second, as we are interested in determining whether the combination of minor mergers and the effects of the SMBH can result in a quenched galaxy similar to what we see in the GM suite, a larger DM-only simulation would not be sufficient due to the lack of baryonic physics.

From R25, we examined 26 MW-mass galaxies, 8 of which are quenched by $z = 0$. To create this sample, we selected all the MW-mass galaxies in R25 with $M_{\mathrm{vir}}$ between $5 \times 10^{11} M_{\odot}$ and $2 \times 10^{12} M_{\odot}$ at $z = 0$ that were not satellites of a more massive halo. There were 26 MW-mass galaxies with these characteristics, each with varying star formation and accretion histories. From each galaxy, we selected every minor satellite merger ($q > 3$) that had a mass ratio between 3 and 20 that occurred within $z = 0.5 - 2$.

We find that 70\% (18/26) of MW-mass galaxies in R25 experience multiple minor mergers occuring within 1 Gyr of each other. Of this population, one galaxy experiences a peak in SMBH activity associated with the merger event which then quenches within a few hundred Myrs, similar to our two quenched GM simulations. This quenched galaxy is one of 8 MW-mass galaxies that are quenched at z = 0 in the simulation. The total number of galaxies in R25 is therefore too small to make a meaningful statistical statement. Nevertheless, the existence of a single example within such a small volume confirms that the minor merger and AGN scenario we have outlined will arise completely naturally and contribute to quenching in LCDM cosmologies. Future analyses based on larger volume simulations can confirm this result.



\section{Summary and Conclusions}
\label{sec-conclude}
Using the genetic modification technique of \citep{Roth2016}, we’ve created a suite of genetically modified galaxies using an initial ``Organic'' MW-mass galaxy with an LMC-mass satellite at $z = 0$. We use the GM process to create galaxies within DM halos with identical large scale structure environments and nearly-identical halo growth histories but for slight variations in their satellite accretion history. The result is a set of four MW-mass halos with accretion histories which have been modified in this controlled way. Despite their overall similarities, we find significant differences in their baryonic evolution. Two of these galaxies remain star forming and two become quenched at $z = 1$. By examining the two quenched cases, we determine that a pair of minor satellite interactions at $z = 1$, concurrent with the peak SMBH accretion rate in the galaxy, can fully quench its star formation until $z = 0$. 

In the two quenched galaxies, the genetic modification process results in a change to the timing of early satellite mergers. Thus, the two satellites interact with the main galaxy within a period of $\sim$ 100 Myrs at $z = 1$. The first satellite merges with the main galaxy, adding to the fuel available to the SMBH, while the second passes through the main galaxy in a flyby. These minor satellite interactions disrupt the disk, and are followed by a peak of SMBH activity within a few hundred Myrs. 

The timing of these events is roughly consistent with the rapid delay times observed between mergers and AGN activity by \cite{Schawinski2014}. They examined GALEX SFRs and SDSS colors of a sample of galaxies from the Galaxy Zoo Citizen science project \citep{Lintott2011} and found that early-type galaxies are quenched by rapid processes ($t_{quench}$ \textless\ 250 Myrs). Other observational results find longer quenching timescales using similar methods \citep[e.g.][]{Schawinski2010, Smethurst2015}.

To better understand our results on a broader scale, we estimate the number of these nearly simultaneous events within a cosmological volume that eventually quench a MW-mass galaxy. In a sample of 26 isolated MW-mass galaxies from the {\sc Romulus25} simulation, there are 10 galaxies that quench by $z=0$ and one experiences multiple minor mergers that coincide with a peak in SMBH accretion that result in a quenched galaxy.

Given current observational capabilities, assessing the impact of minor mergers on star formation history remains a challenge. However, work disentangling minor merger effects on SF galaxies has been ongoing \citep[][and references therein]{Maschmann2020} and JWST will likely improve upon these observations in the future.

Major mergers ($q < 3$) between massive galaxies have long been thought to be the primary means by which spiral galaxies transform into ellipticals \citep{DiMatteo2005,Springel2005,Hopkins2006,Somerville2008,Johansson2009,Schawinski2010}. In addition, these mergers drive starbursts and fuel central SMBHs, where the latter process may suppress star formation in the remnant galaxy \citep{Richards2006,Reddy2008,Hopkins2010,Sanchez2017}. The tidal torques, combined with the angular momentum of infalling gas,  funnel gas into the center of the galaxy, which subsequently increases the accretion rate of the SMBH \citep{Barnes1996, DOnghia2006, Hopkins2009}. However, recent observational and theoretical studies have called into question the efficacy of major mergers in driving SMBH fueling \citep{Fanidakis2012, Hirschmann2012, DelMoro2016, Hewlett2017, Villforth2018, Steinborn2018}. For example, \cite{DelMoro2016} examined a sample of luminous mid-IR quasars and found no direct evidence linking SFRs and AGN luminosity. 

While previous work has explored the galaxy-scale physical consequences of major mergers, the role of minor merger disruption in galactic evolution and SMBH fueling is less understood \citep[but see:][]{Toomre1972, Ostriker1980, Carlberg1986, Kormendy1989, Hopkins2009}. A recent simulation study by \cite{Hani2020} explores the relationship between mergers and galaxy evolution. They find that both major ($q \lesssim 3$) and minor mergers ($q \gtrsim 3$) can significantly increase the sSFR of the post-merger galaxy. However, the enhancement of the sSFR is a factor of $\sim$ 2 for minor mergers ($q \sim 3-10$) and $\sim 2.5$ for major mergers. While \cite{Hani2020} do not find that galaxy mergers are globally quenching their post-merger galaxies, they conclude that the strongest merger-driven galaxies become quenched faster than their control galaxies.

We therefore explore the role of minor mergers, in tandem with the feedback of the SMBH, as drivers for quenching massive galaxies. The closely timed interaction of the minor merger and flyby ultimately disrupt the galaxy disk, and drive gas into the vicinity of the SMBH, thereby fueling it \citep{Kormendy2013}. Thus, it is the sequence and combination of these events that occur during a short period of few hundred Myrs \textemdash the SMBH fueling and subsequent feedback coupled with the disruption of the disk \textemdash that fully quench both MW-mass galaxies in GM2 and GM3 by $z = 1$. Our study has revealed a complex story where the dual impact of two minor mergers, and the increased SMBH fueling that these mergers drive, create a viable pathway for quenching in MW-mass galaxies and supports the growing evidence that the mechanisms that quench a galaxy are numerous and varied. \\


\section{Acknowledgements}
This work was supported by the FINESST19-23 grant 80NSSC19K1409. NNS gratefully acknowledges helpful conversations with Alyson Brooks, Ferah Munshi, Jillian Bellovary, Iryna Butsky, and Samantha Benincasa. The authors thank the referee for their thoughtful comments and suggestions in improving the initial draft of this paper. This work used the Extreme Science and Engineering Discovery Environment (XSEDE), which is supported by National Science Foundation grant number ACI-1548562. The work used the XSEDE supported Stampede2 at TACC through allocation TG-MCA94P018. Resources supporting this work were provided by the NASA High-End Computing (HEC) Program through the NASA Advanced Supercomputing (NAS) Division at Ames Research Center. MT gratefully acknowledges support from the YCAA Prize Postdoctoral Fellowship. This project has received funding from the European Union's Horizon 2020 research and innovation program under grant agreement No. 818085 GMGalaxies. AP was further supported by the Royal Society. These results were obtained using the publicly available analysis softwares: pynbody \citep{pynbody}, TANGOS \citep{TANGOS}, yt \citep{Turk2011a}, numpy \citep{harris2020array}, matplotlib \citep{Hunter:2007}, python \citep{CS-R9526}, pandas \citep{reback2020pandas}. The authors acknowledge Paul Tol's detailed notes on creating accessible color schemes for scientific figures (https://personal.sron.nl/~pault/).
\newpage
\bibliography{./SanchezH2QaG22019.bib}

\end{document}